\documentclass[aip,amsmath,amssymb,reprint,floatfix,nofootinbib]{revtex4-2}
\usepackage{graphicx}
\usepackage{hyperref}

\begin{document}
\title{The Motion of Curling Rocks: Experimental Investigation and Semi-phenomenological Description}
\author{E.T. Jensen}
\email[email:]{ejensen@unbc.ca}
\author{Mark R.A. Shegelski}
\affiliation{Department of Physics \\ University of Northern British Columbia \\ 3333 University Way, Prince George B.C. V2N 4Z9, Canada}

\date{February 27, 2004}

\begin{abstract}
A large number of curling shots using a wide range of rotational and translational velocities on two different ice surfaces have been recorded and analyzed. The observed curling rock trajectories are described in terms of a semi-phenomenological model. The data are found to rule out `dry friction' models for the observed motion, and to support the idea that the curling rock rides upon a thin liquid film created at the ice surface (i.e. `wet friction'). Evidence is found to support the hypothesis that the frictional force acting upon each segment of the curling rock is directed opposite to the motion relative to this thin liquid film and not relative to the underlying fixed ice surface.
\end{abstract}
\maketitle

\section{Introduction}
In the sport of curling, cylindrically symmetrical granite rocks, called curling rocks, are projected over a sheet of ice with a small initial angular speed.  There are several aspects of the motion of curling rocks that are of great physical interest.  Some examples are as follows:
\begin{itemize}
\setlength\itemsep{0em}
\item The trajectory of a curling rock is not a straight line but is instead curved (hence the name ``curling'').
\item The direction of the lateral motion, the ``curl'', on ice is opposite to that of a geometrically similar cylinder sliding over a smooth dry surface.
\item The net lateral displacement, or curl,  is large enough that it is not easily understood.
\item The dependence of the net curl on the initial angular speed is weak, which is also not readily understood.
\end{itemize}

There has recently been an increased interest in the underlying physics of motions of curling rocks, mostly of a theoretical nature. In order to obtain a better idea of what theoretical ideas could serve as part of a fully satisfactory model, we have set ourselves the task of collecting a large amount of experimental data.  We examine a variety of cases, ranging from the rock traveling  a long distance (about 25m) and rotating only 2 or 3 full rotations, to a short travel distance of up to a few metres with a large number of full rotations (as many as about 80), to pure rotation with the number of full rotations going from roughly 1 to 80.  We also used different ice surfaces: {\em pebbled} ice, and {\em flooded} ice.

\section{Review of Previous Work}
Two aspects of curling should be briefly explained at this point. First, the bottom of the curling rock is not flat, but is curved and hollowed such that the rock rides upon a flat contact annulus made of granite. Second, the ice surface is not flat-- it consists of mm-sized ice mounds called ``pebbles", and is referred to as ``pebbled ice". The details of the ice preparation and curling rock are discussed more fully in Section {\ref{Experimental_Details}}. The question as to the underlying physical reasons for the observed motion of curling rocks has, perhaps surprisingly, only recently been addressed in a physically reasonable manner. 

Johnson presented a physical model that was based on an asymmetry in the friction around the contact annulus \cite{Johnson:1981}.  The physical reason for asymmetrical friction was taken to be due to heating of the ice by the rock.  That the normal force on the front half (i.e. the leading half) of the contact annulus exceeds the normal force on the back (trailing) half led to taking the ice at the front as being heated to a higher temperature than that at the back.  This led in turn to the coefficient of friction being smaller at the front as compared to the back, thus accounting qualitatively for the direction of the lateral motion.  Melting of the ice was not taken into account in this model.

Some time later, our group proposed a model that took into account the kinetic melting of the ice to give a thin liquid film nested between portions of the contact annulus  and the underlying solid ice \cite{Shegelski:1996}.  The model used a velocity-dependent coefficient of friction, with the friction having a front-back asymmetry,  being less at the front than at the back, a consequence of more melting at the front as compared to the back due to the normal force being larger at the front half than at the back.  The coefficient of friction was taken to be of the form $\mu=\alpha v^{\phi}$.  The value of $\phi$ that we used in our first paper has been supplanted by a better choice that emerged upon subsequent research; we discussed this in a follow-up paper \cite{Shegelski:2003} and we will return to this later in this paper. 

In several other papers, we investigated various aspects of the motions exhibited by curling rocks and other sliding, rotating cylinders.  We predicted and observed that rapidly rotating curling rocks that slide slowing over pebbled ice can stop sliding long before they stop rotating \cite{Shegelski:1999-1}.  A key ingredient in this work was the thin liquid film between the annulus and the ice.  We then predicted and observed that a cylinder with a different contact pattern would stop rotating before it stopped sliding on pebbled ice \cite{Shegelski:1999-2}.  We also calculated trajectories of rotating cylinders sliding over a smooth surface for which there was no melting and for which the coefficient of friction was constant \cite{Shegelski:1999-3}.  We presented exact equations and analytical solutions.  Our results were subsequently further confirmed by some numerical and experimental results reported by Penner \cite{Penner:2001}.  We also presented a simplified version of our model, using a constant coefficient of friction, for convenience \cite{Shegelski:2000}, and we presented exact equations and analytical results for trajectories of cylinders with extreme geometries \cite{Shegelski:2000, Shegelski:2001}.  At one extreme was cylinders with small contact rings and large moments of inertia \cite{Shegelski:2001};  at the other extreme were small moments of inertia and large contact rings \cite{Shegelski:2000}.  We showed, among other things, that one can have sliding and rotation stopping at different times due to the extreme geometries.  We also investigated the motion of {\em extremely} rapidly rotating cylinders and showed that large lateral deflections can be obtained \cite{Shegelski:2002}.  This is addressed later in this paper.

Denny has presented two different approaches. In one case, an attempt was made to describe the lateral motion as being due to a torque, and the formulation was made in non-inertial reference frames \cite{Denny:1998}.  This approach was shown to be unacceptable on physical grounds \cite{Shegelski:1999-3, Penner:2001}.  More recently, Denny proposed that the curl is due to a front-back asymmetry \cite{Denny:2002}, as was first proposed by Johnson \cite{Johnson:1981} and later by our group \cite{Shegelski:1996,Shegelski:2003,Shegelski:1999-1,Shegelski:1999-2,Shegelski:1999-3,Shegelski:2000, Shegelski:2001, Shegelski:2002}.  In this more recent model, Denny suggested the asymmetry to be due to the rock grinding the ice, resulting in ice granules lodged into portions of the roughened contact ring, leading to part of the ring moving with ``ice-on-ice" friction and part with granite on ice friction.  The proposal was that the friction would be smaller at the front than at the back because the front would have ice-on-ice friction and the back mostly granite on ice friction. However, the coefficient of kinetic friction for ice-on-ice is about 0.03 \cite{Oksanen:1982}, which is about 3 times larger than the coefficient of friction for granite on ice (which is of order $0.01$). The granular model also attempted to account for the weak dependence of the total lateral deflection on the initial angular speed $\omega_0$.  This was done by taking the effect of the granules to act over a portion of the contact annulus  that changes with increasing $\omega_0$ in just the manner needed to give the weak dependence on $\omega_0$.  This seems to be somewhat artificial and lacks a convincing physical basis.  Of note is that the granular model assumes dry friction (i.e. constant $\mu$) \cite{Denny:2002}; we will discuss the implications of this later in this paper.

Experimental results from Penner \cite{Penner:2001} showed that the net lateral deflection depends weakly on the initial angular speed, or the total number of rotations of the rock.  We report results herein that agree with and go beyond those of Penner's.  Penner also reported experimental work concerning the value of the exponent $\phi$; we also address this point. Although it was hypothesized in \cite{Penner:2001} that stick-slip friction can account for observed motions of curling rocks, no theoretical treatment was given.

The experimental results reported in this paper show that  the thin liquid film model captures many qualitative features of the observed motions, and also requires modifications to fully describe the observed motions.  Denny's granular model, based on dry friction, cannot account for the experimental results reported here.  We discuss this at the end of this paper.

\section{Background Theory}
We next present the formulae that are used in our analysis of the experimental data. These formulae are based on semi-phenomenological considerations as described below. We haved categorized the different curling shots as follows: pure rotation; rapid rotation, slowly sliding (RRSS); and draw shots. 

On theoretical grounds \cite{Penner:2001,Oksanen:1982,Persson:2000,Evans:1976,Akkok:1987}, we assume that the coefficient of
kinetic friction has the form $\mu=\alpha [u(\theta,t)]^{\phi}$ where $\alpha$
and $\phi$ are constants and $u(\theta,t)$ is the speed at time $t$ of the
portion of the contact annulus located between $\theta$ and $\theta+d\theta$.

\subsection{Pure Rotation}
In the case of pure rotation, we immediately have 
\begin{equation}
\tau = -r \alpha [r\omega(t)]^{\phi} Mg = I d\omega(t)/dt
\end{equation}
where $I$ is the moment of inertia of the rock: $I\approx\frac{1}{2}MR^2$.
This gives
\begin{equation}
\omega(t) = \omega_0 \left( 1- \frac{t}{t_{F_\theta}} \right)^{1/(1-\phi)}
\label{w(t)_Eqn}
\end{equation}
and
\begin{equation}
\theta(t) = \theta_F \left[ 1 - \left( 1- \frac{t}{t_{F_\theta}} \right)^{\lambda_{\theta}} \right]
\label{q(t)_Eqn}
\end{equation}
where the stopping time for rotation is given by
\begin{equation}
t_{F_\theta}=\frac{ \omega_0^{1-\phi} R^2 }{ 2 (1-\phi)r^{1+\phi} \alpha g}
\label{rot_stop_time}
\end{equation}
and
$\theta_F = \frac{1-\phi}{2-\phi} \omega_0 t_{F_\theta} $ and $\lambda_{\theta}=\frac{2-\phi}{1-\phi}$. Comparing the theoretical and observed quantities gives $\phi = \frac{ 2 - \lambda_{\theta} }{ 1-\lambda_{\theta} }$ and
$\omega_0 = \frac{ \theta_F \lambda_{\theta} }{t_{F_\theta}}$
which in turn gives
$\alpha = \frac{ \omega_0^{1-\phi} R^2 }{ 2(1-\phi)r^{1+\phi} g t_{F_\theta} }$. We have used Eq. {\ref{q(t)_Eqn}} extensively in fitting observed curling rock rotation data; in this context the equation is used such that $t_{F_{\theta}}$ and the $(\theta_i,t_i)$ are measured from experiment, and $\theta_F$ and $\lambda_{\theta}$ are parameters obtained by a Marquandt non-linear fitting  procedure.

The physical meaning of the $\lambda_{\theta}$ exponent is that for larger values of $\lambda_{\theta}$, the rotational motion ceases more rapidly, i.e. the friction is larger. This formula for rotation is used in analyzing the experimental data presented in this paper. Although Eq. {\ref{q(t)_Eqn}} was derived from consideration of the case of pure rotation, we have found that this expression also appears to reasonably describe rotational motion for the cases of rotational-dominated ($r\omega>>v_{translation}$) and translation-dominated ($r\omega<<v_{translation}$) motion. These results will be described in detail in section \ref{Observations}.

\subsection {Rapid Rotation, Slow Sliding}
In the case of rapid-rotation--slow-sliding (RRSS), we anticipate that
the rotation will be of the form given in Eq. {\ref{q(t)_Eqn}} above, and further that the value of $\lambda_{\theta}$ will be close to the value for pure rotation. 
Observed motions discussed below support this.

We also try to fit the translation using formulae for $y(t)$ and $x(t)$ that are derived in the next subsection.

\subsection { Draw Shot }
The draw shot in the sport of curling has the feature that the 
rotational speed, $r\omega$, is much less than the translational 
speed $v$.

The dominant motion here is displacement in the original direction of 
motion.  To leading order, the equation for this motion is
\begin{equation}
\vec{F}_v = - \alpha [v(t)]^{\phi} Mg \hat{e}_v
\label{Fv_Eqn}
\end{equation}
where $\hat{e}_v$ is a unit vector in the instantaneous direction of motion. This gives
\begin{equation}
v_y(t) = v_0 \left( 1 - \frac{t}{t_{F_S}} \right)^{1/(1-\phi)}
\label{vy_Eqn}
\end{equation}
and
\begin{equation}
y(t) = y_F \left[ 1 - \left( 1- \frac{t}{t_{F_S}} \right)^{\lambda_y} \right]
\label{y(t)_Eqn}
\end{equation}

and
\begin{equation}
t_{F_S} = \frac{ v_0^{1-\phi} }{ (1-\phi)\alpha g }
\label{tFs_Eqn}
\end{equation}
where $y_F = \frac{1-\phi}{2-\phi} v_0 t_{F_S}$ and $\lambda_y=\frac{2-\phi}{1-\phi}$. Comparing the theoretical and observed quantities gives $\phi = \frac{ 2-\lambda_y }{ 1-\lambda_y }$ and $v_0 = \frac{ y_F \lambda_{y} }{t_{F_S}}$ which in turn gives $\alpha = \frac{ v_0^{1-\phi} }{ (1-\phi) g t_{F_S} }$.

We have used Eq. {\ref{y(t)_Eqn}} as a fitting function for $y$-displacement data from various curling shot data. For this application, the $t_{F_{S}}$ and $(y_i,t_i)$ are obtained from the data and the parameters $y_F$ and $\lambda_y$ are obtained from a Marquandt non-linear fitting procedure.

Of great interest for curling is the motion lateral to the original direction, 
i.e. the displacement $x(t)$.  In terms of the components of the force 
opposite to the instantaneous direction of motion, $\vec{F}_v$, and 
transverse to the instantaneous direction of motion, $\vec{F}_T$, the 
component of the force lateral to the initial direction of motion, $\vec{F}_x$,
is given by
\begin{equation}
 \vec{F}_x = \vec{F}_T \cos[\psi(t)] + \vec{F}_v \sin[\psi(t)] 
\end{equation}
where $\psi(t)$ is the angle between the initial direction of motion 
and the instantaneous direction of motion, and is obtained by solving
\begin{equation}
 M \Omega(t) = \frac {F_T(t)} {v(t)}
\end{equation}
for $\Omega(t)$ and then solving

\begin{equation}
 \psi(t) = \int_0^t \Omega(t')  dt'
\end{equation}
For $\vec{F}_v$ and $v(t)$ we use the expressions for the 
displacement $y(t)$.  We thus require only an expression for $F_T(t)$.  From analysis of the data (discussed below), we were led to consider the case of $F_T(t)$ approximately  constant. Then, since $\psi(t)$ is small for the draw shot, we have, to a very good approximation, that the acceleration $a_x$ in the $x$-direction is given by
\begin{eqnarray}
a_x(t) &=& a_T - |a_v(t)| \; |\psi(t)|  \nonumber \\
 & =& a_T - a_T\frac{1}{\phi}\left[1- \left(1-\frac{t}{t_{F_S}}\right)^{\phi/(1-\phi)}\right]
\end{eqnarray}
where $a_T$ is a constant.  Integrating gives the expression for the 
lateral deflection, i.e. the curl:
\begin{eqnarray}
 x(t) & = & \frac{1}{2}a_T t_{F_s}^2 \left( \frac{t^2}{t_{F_S}^2} - \frac{1-\lambda_y}{2-\lambda_y}\left[ \frac{t^2}{t_{F_S}^2}+ \frac{2}{1-\lambda_y}\left(\vphantom{\frac{1}{1}} \right. \right. \right. \nonumber \\
& &  \left. \left. \left. 
 \frac{t}{t_{F_S}} - \frac{1}{\lambda_y}
\left[1-\left(1-\frac{t}{t_{F_S}}\right)^{\lambda_y}
\right]
\right)
\right]
\right)
\label{x(t)_Eqn}
\end{eqnarray}
where the total $x$-displacement is $x_F=\frac{1}{2}\frac{1}{\lambda_y} a_T t_{F_S}^2$. We have used Eq. {\ref{x(t)_Eqn}} as a fitting function for experimentally observed $x$-displacements. Note that we use $\lambda_y$ because the draw shot is 
translation-dominated. If we assume that the $\lambda_y$ parameter is fixed and determined from $y(t)$ displacement data, then fitting $x(t)$ displacement data using Eq. {\ref{x(t)_Eqn}} is a simple one-parameter fit to obtain $a_T$. If we relax the fitting constraint to allow for $\lambda$ different from $\lambda_y$, then we have a two parameter non-linear fit to obtain $a_T$ and $\lambda$.

For the rotation of the rock for a draw shot, we again have used Eq. {\ref{q(t)_Eqn}}, as we have found that this equation properly represents the observed rotational motion.

There are several aspects of the draw shot that are 
intriguing.  Two of the most puzzling features are (a) the magnitude of
the lateral motion and (b) the weak dependence of the lateral 
deflection on the initial angular speed.

\subsection{Implications of Wet Friction}
In previous treatments of the kinetic friction of surfaces on ice \cite{Evans:1976,Akkok:1987}, including one dealing with curling rock motion \cite{Penner:2001}, a form for the kinetic friction of $\mu_k\propto v^{-1/2}$ has been used.  This form with $\phi=-1/2$ results by combining the equation for conduction of heat from the rock into the ice with the equation for the amount of energy transfered.  The heat conduction equation gives the rate of transfer of heat:
\begin{equation}
\frac{\Delta Q}{\Delta t} = kA \frac{\Delta T}{\Delta z}
\label{DQ_Eqn1}
\end{equation}
where $k$ is the thermal conductivity of the ice, $A$ is the area of the ice in the $x$-$y$ plane that is in contact with a portion of the contact annulus of the rock, $\Delta T$ is the difference in temperature between the ice in contact with the portion of the rock (i.e. $T=0^{\circ}$C), and the ``bulk'' temperature (i.e. $T=-5^{\circ}$C) at a depth $\Delta z$ under the surface.
The amount of energy transferred from the rock to the ice can be written in two different ways, one of which is
\begin{equation}
\Delta Q = \rho A \Delta z (c\Delta T + l)
\label{DQ_Eqn2}
\end{equation}
where $\rho$ is the density of ice, $c$ is the specific heat, and $l$ is the latent heat of melting.
One obtains $\phi=-1/2$ by solving Eq. {\ref{DQ_Eqn1}} for $\Delta z$ and putting the result into Eq. {\ref{DQ_Eqn2}} to solve for $\Delta Q$ to get:
\begin{equation}\Delta Q = \left( k \rho (c\Delta T + l) \Delta T A^2 \Delta y  \right)^{1/2}  v^{-1/2}
\end{equation}
where $\Delta y=v\Delta t$ is the distance that the rock slides in time $\Delta t$. The final step is to write the energy transferred from the rock to this piece of ice as the rock slides a distance $\Delta y$:

\begin{equation}
\Delta E = \mu f Mg \Delta y
\end{equation}
where $f$ is the fraction of contact area $A$ compared to the total contact area of the annulus with the ice.

Combining the last two equations give $\mu = \alpha v^{-1/2}$.  That we do not observe $\phi=-1/2$ indicates that one or more assumption in the derivation is incorrect.  Since the value $\phi=-1/2$ comes from the $\Delta t$ in the heat conduction equation, it would seem that use of this equation should be questioned. Indeed, a first principles calculation for viscous drag upon a flat rotating cylinder with a thin fluid interface by Montgomery \cite{Summers:2003} gives $\lambda_{\theta}=2.95\pm0.15$ (i.e. $\phi=+0.49$, obtained by fitting the calculated theoretical data points to Eq. {\ref{w(t)_Eqn}}), which seems to us to be an additional motivation for a fresh approach to calculating $\alpha$ and $\phi$.

In our view, due to the existence of a thin film between contact points and the solid ice, it is feasible that the direction of the friction could be changed to be in the direction of the motion of the contact portion relative to the immediate thin liquid film instead of relative to the underlying solid ice. Our motivation for this is as follows. If the direction of friction is opposite to the direction of motion of a portion of the contact annulus relative to the underlying solid ice, then the calculated curl distance is too small to agree with observed curl distances. Moreover, rotational and translational motions would cease almost simultaneously for rapid rotation, slow sliding curling rocks, which is not what is observed (e.g. see Section {\ref{RRSS_Section}}). Since the draw shot motion is dominated by translation, we can model this change in direction of the frictional force by altering the semi-phenominological equation of motion for rotation by the introduction of a new constant that represents the change in direction of the force of friction. Thus for rotation of the rock for a draw shot, we find that starting with the leading order term for the torque \cite{Shegelski:2000}
\begin{equation}
\tau_z(t) = -\frac{1}{2}C_{\omega} r\alpha v^{\phi}(t) Mg 
\frac{r\omega(t)}{v(t)} = \frac{1}{2} M R^2
\end{equation}
where $C_{\omega}$ is a constant to be determined, gives an expression for $\omega(t)$ that is in accord with our experimental results, namely

\begin{equation}
\omega(t) = \omega_0 \left( 1- \frac{t}{t_{F_{\theta}}} \right)^{\lambda_{\theta}-1}
\end{equation}
where 
\begin{equation}
C_{\omega} = \frac{\lambda_{\theta}-1}{\alpha g v_0^{\phi-1}\left( r/R \right)^2t_{F}}=\left( \frac{\lambda_{\theta}-1}{\lambda_y -1}\right) \left( \frac{R}{r}\right)^2
\label{Cw_Eqn}
\end{equation}
If one assumes that the direction of the friction on a portion of the contact annulus is in the direction opposite to the direction of motion of that part of the annulus relative to the underlying solid ice (and {\em not} relative to the thin liquid film), then one finds that the calculated lateral deflection of the rock is somewhat less than the actual deflection. If instead the direction is opposite to that of the motion relative to the thin film, the calculated value of the lateral deflection agrees with the observed value. If this is the case, then we would expect $C_{\omega}$ for a draw shot to be of order 1 but not equal to 1.

For the case of a rapidly-rotating slowly sliding rock, it is more appropriate to consider the translation as a perturbation on the dominant rotational motion. The leading order term for $F_v$ is \cite{Shegelski:2001}
\begin{equation}
F_v=-\frac{1}{2}C_{y}\alpha (r\omega(t))^{\phi}Mg\frac{v(t)}{r\omega(t)}
\end{equation}
which includes an additional factor $C_y$ to be determined. This results in an expression for $v(t)$ of the form
\begin{equation}
v(t)=v_0 \left(1-\frac{t}{t_{F_s}}\right)^{\lambda_y -1}
\end{equation}
where
\begin{equation}
C_y=\frac{2(\lambda_y -1)}{\alpha g (r\omega_0)^{\phi-1}t_{F_{\theta}}}= 4 \left(\frac{r}{R}\right)^2 \left(\frac{\lambda_y -1}{\lambda_{\theta}-1}\right)
\label{Cy_Eqn}
\end{equation}
If the frictional force acting upon the rock were in the direction opposite that of the instantaneous velocity vector relative to the ice, then one would expect that $C_y=1$; we anticipate that the value observed for RRSS shots should have $C_y$ of order 1 but not equal to 1.

Another consequence of having the frictional force acting upon a piece of the curling rock being in the direction relative to the thin liquid film rather than the underlying ice is that the rotation and translational motions of a curling rock need not cease simultaneously \cite{Shegelski:2003}. As we discuss in Section {\ref{Observations}}, there are cases where this is clearly observed. We thus anticipate finding $C_{\omega}\ne 1$ and $C_y\ne 1$. Our semi-phenomenological model will involve the following parameters that need to be taken from observations:  $\alpha$, $\lambda_{\theta}$, $\lambda_y$, $a_T$, $C_y$ and $C_{\omega}$.

\section{Experimental Details}
\label{Experimental_Details}
\subsection{Ice Surfaces}
The condition of the ice surface upon which the game of curling is played has a dramatic effect upon the curling rock motion. For a standard curling rink, a nominally flat ice surface at $-5^{\circ}C$ is {\it pebbled\/} by spraying droplets of water over the ice surface so that the droplets freeze promptly to form a surface containing a myriad of mm-sized bumps. There is considerable art in preparing a good curling rink surface. In the present work we used the facilities of the Prince George Golf and Curling Club and their preparation expertise. Since the ice surface condition is so crucial in the motion of curling rocks, we have explored curling rock motion on a few other ice surface preparations. We also present some data using an ice surface that is obtained after flooding the rink and freezing this water, which we call {\it flooded ice\/}. In the flooded ice we have used, the ice surface is quite flat but is not perfectly smooth. The flooded ice surface has considerable small scale roughness, which forms upon the surface during freezing. As will be discussed below, this roughness leads to rather large curl of the rocks compared to standard pebbled ice, but is more fragile than pebbled ice due to the small size of the roughness features, which wear down rapidly. In order to stress the role of the ice surface on curling rock motion, we also mention here that we have qualitatively observed curling rock motion on two other ice surface preparations, {\it burned\/} ice and {\it scraped\/} ice. Scraped ice results from the use a mechanical scraping blade on pebbled ice (this is done routinely to resurface a curling rink), and leaves an ice surface with considerable roughness over a wide range of length scales. Burned ice is obtained after using a torch to remelt the scraped ice surface layer which then refreezes promptly. In this case, the ice surface is glassy smooth without significant surface roughness. Curling rocks thrown on scraped ice exhibit very large curl distances. Curling rocks thrown on burned ice travel along a near-perfect straight trajectory and exhibit little or no curl at all. 

\subsection{The Curling Rock}
A standard curling rock is disc shaped with its approximate dimensions being 29cm in diameter, 11.4cm in height and 20.0kg in mass. At the present, curling rocks are all made from dense granite, most of which is quarried in Scotland or Wales. The bottom surface of the curling rock is dished, and the rock rides on the ice on a narrow running band, which has a diameter of approximately 12.5 cm and a width of 4 to 5 mm. Given the pebbling of the ice surface, only a small percentage (roughly 5\%) of the running band is in contact with the ice at any given moment. The running band on a standard rock is prepared by a sequence of abrasions so as to increase the friction between the granite and the ice, following procedures that vary between manufacturers. A diagram of a representative curling rock is presented in Fig. 7 of Ref. \cite{Penner:2001}.

It is worth noting that the interface between the curling rock running band and the ice plays a crucial role in the observed lateral motion. Curling stones that have a polished granite running band exhibit little or no curl. As another example, we have observed that a rock fitted with a polished {\em stainless steel} running band exhibited {\em larger} curl than a regular curling rock, emphasizing that the interface characteristics play a large role in determining the motion and also that a large curl distance on pebbled ice does {\em not} require that the running band be roughened. We note that the same qualitative curling motion is observed for the roughened granite contact annulus as for the smooth stainless steel annulus. It is therefore reasonable to attribute the curl in both cases to a common mechanism which we consider to be due to a thin liquid film. We also stress that the results discussed in the present work are taken with one particular curling rock in order to remove variations due to differently prepared rocks. Thus it is likely that other curling rocks thrown on different ice surfaces would exhibit somewhat different motions than those reported here.

\subsection{Data Collection}
One of the goals of the current work was to record curling rock trajectories under a range of ice conditions and for a variety of different curling rock initial conditions. These trajectories were recorded using a small digital video camera. For the majority of curling rock shots, the video camera was mounted on a boom looking vertically downward, and the boom was suspended atop a stepladder. This stepladder was slid down the ice rink by an attendant, paralleling the curling rock as it moved down the curling rink. The ladder speed and position was adjusted so that the camera field of view always included the curling rock. For the low initial velocity curling shots such as pure rotational and rapidly rotating, slowly sliding shots, the video camera was sometimes simply handheld at shoulder height above the curling rock. The video camera we used recorded at a rate of 30 frames per second, which limited our time resolution to $\pm0.03$ s.

The curling rock position was determined using a grid from which the $x(t)$ and $y(t)$ coordinates of the curling rock could be measured.  The grid was created by suspending fine coloured thread just under the ice surface. The curling rock rotation was monitored by monitoring the position of coloured markers affixed to the top of the curling rock. Since the video camera orientation was fixed throughout the motion, measurement of the instantaneous angle of the curling rock was straightforward.

A large number of curling shots were recorded and analyzed in order to be sure that the trajectories reported here would be reasonably representative. For a number of recorded trajectories, the rocks would exhibit unusual motions, such as wobbling back and forth and sudden changes in rotation rate or $x$-velocity. These anomalous shots could be caused by a number of different factors, such as picking up some grit during the motion, striking unusually large ice pebbles or moving in a patch of the rink that is significantly tilted or dished. We removed these shots from consideration when appropriate.

\subsection{Data Analysis}
The measurements of curling rock trajectories were obtained by downloading the digital video to a Macintosh computer workstation, and the video was then analyzed frame-by-frame. The shot start time $t=0s$ was picked for a video frame just after the rock was released by the curler, and the shot stopping time was picked from the frame for which no additional motion could be distinguished from the prior frame. We have found that in some instances there are different stopping times for translational motion and for the rotational motion-- in the present work we use the notation $t_{F_S}$ and $t_{F_{\theta}}$ respectively to denote these times.

The grid system was chosen with the $y$-axis along the long axis of the rink and the $x$-axis perpendicular to that. Since there were very large differences between the $x$- and $y$-velocities for the rocks, the $x(t)$, $y(t)$ and $\theta (t)$ motions were recorded separately. Although the curlers attempted to throw the curling rocks with zero $x$-velocity, in some cases the shots had small non-zero initial $x$-velocities. Since the theory assumed zero $x$-velocity, these skew shots could only be analyzed by rotating the data points such that the initial $x$-velocity was zero.  In practice, for the curling shots to be discussed in the present work, the rotation angle was zero in many cases or very small ($<2.5^\circ$).

\section{Observations and Discussion}
\label{Observations}
\subsection{Pure Rotational Motion}
\begin{figure}
\caption{Curling rock undergoing pure rotation on pebbled ice, with a fit to Eq. {\ref{q(t)_Eqn}} shown with residuals. The fit to the data points is so close that the points cannot be distinguished from the fitting line drawn. In this data, the rock rotated for a total of 10.4 turns.}
\includegraphics[scale=0.8]{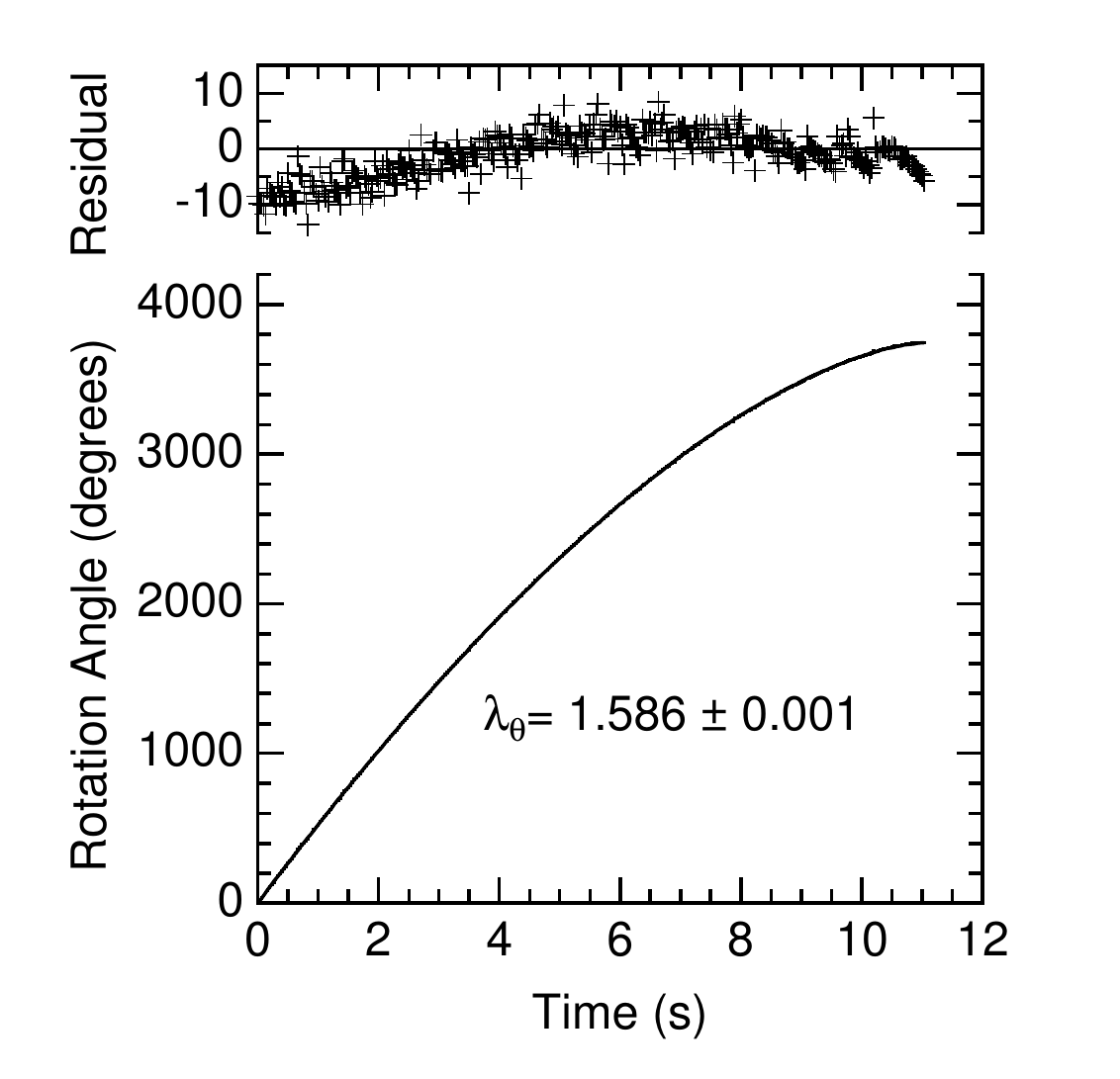}
\label{PR_PI_Data1}
\end{figure}

\begin{figure}
\caption{Plot of the extracted $\lambda_{\theta}$ from a number of cases of curling rocks undergoing pure rotation on pebbled ice. The $\lambda_{\theta}$ values are largely insensitive to the initial rotation rate of the rock, though there is some indication for slightly smaller values for very slow initial rotation rates.}
\includegraphics[scale=0.75]{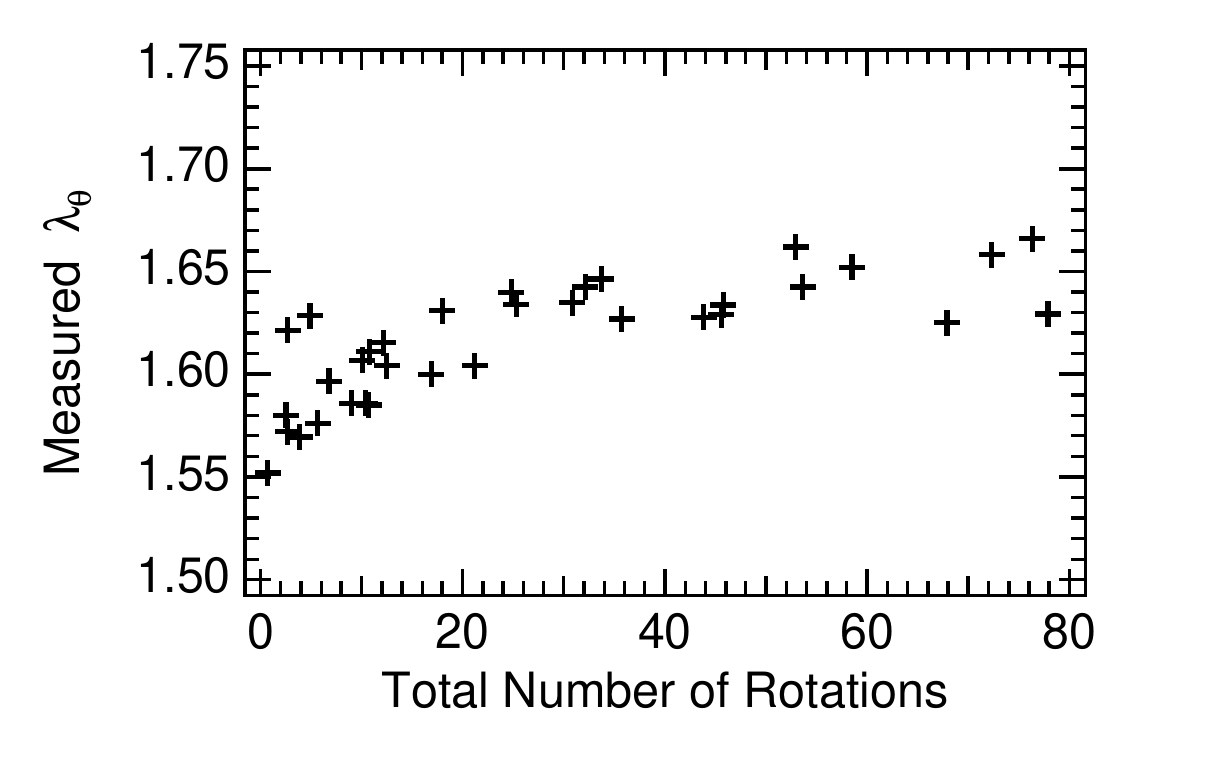}
\label{RotationLambdas}
\end{figure}

Figure {\ref{PR_PI_Data1}} shows data obtained from a curling rock undergoing only rotational motion on standard pebbled ice. In this data, the curling rock makes a total of of 10.4 rotations before stopping. This data was obtained using a custom computer program to record the curling rock position angle from each frame of the recorded video. This data was found to be very well represented by fitting to Eq. {\ref{q(t)_Eqn}}, with the exponent $\lambda_{\theta}=1.586\pm0.001$. We examined a large number of such pure rotational motions for curling rocks having total rotations from less than 1 total turn to 78 total turns and found that the $\lambda_{\theta}$ exponent varies over a range of $1.55<\lambda_{\theta}<1.66$, but is relatively insensitive to the initial rotational velocity, as shown in Fig. {\ref{RotationLambdas}}. It is possible that there is a small correlation between $\lambda_{\theta}$ and the initial angular speed, as the trials at low initial rotational speed tend to show a somewhat smaller value of $\lambda_{\theta}$. This may be due to the difference in the amount of wear that occurs on the pebbled ice during rotation. Curling rocks that rotate for a long time have more opportunity to wear a groove in the pebbling in the contact ring, and increase the contact area between the ice and the rock. This could increase the friction for the latter part of the motion and result in a larger value of $\lambda_{\theta}$.

The values of $\lambda_{\theta}$ observed in Fig. {\ref{RotationLambdas}} differ considerably from what would be expected if `dry friction' was responsible for slowing the curling rock rotation. In dry friction, the coefficient of kinetic friction $\mu_k$ is velocity independent, corresponding to $\phi=0$ and $\lambda_{\theta}=2$. Clearly, whatever the dominant mechanism for rotational friction is, it is distinct from dry friction.

For $\lambda_{\theta}=1.586 \pm 0.001$, we find $\phi_{rot}=-0.707 \pm 0.003$, and for  $1.55\leq\lambda_{\theta}\leq 1.66$, we find the range of $\phi_{rot}$ to be $-0.82\leq\phi_{rot}\leq -0.52$.  This range is close to but also differs from the theoretical value $\phi=-1/2$ suggested by Penner \cite{Penner:2001}.  We also find
$\alpha_{rot}=0.0094$ (m/s)$^{-\phi_{rot} }$ for Fig. {\ref{PR_PI_Data1}}, which is again close to but different than the value $\alpha=0.0080$ (m/s)$^{1/2}$ presented by Penner \cite{Penner:2001}.

\subsection{Rapidly Rotating Slowly Sliding Motion}
\label{RRSS_Section}
In these shots, a small translational velocity is added to the initial rotational velocity.
For each shot there are three observed data sets: $\theta(t)$, $y(t)$ and $x(t)$. We have used Eq. {\ref{q(t)_Eqn}} as the fitting function for $\theta (t)$ since we expect that the rotational motion here is not significantly different from that of the pure rotation case. We have also attempted to use a fitting function for $y(t)$ of the form given by Eq. \ref{y(t)_Eqn} to determine if it can properly model the $y(t)$ motion.

Note that the parameter $t_{F_S}$, which refers to the observed moment that the translational motion ceases, does not need to coincide with $t_{F_{\theta}}$ (the moment that rotational motion is observed to cease). As we discuss below, we do observe that on pebbled ice, the translational motion can cease before rotation. On the flooded ice surface, we find that $t_{F_S}=t_{F_{\theta}}$ for RRSS shots, within the error of our measurements.

For the motion transverse to the initial direction of motion, $x(t)$, we have attempted to fit the data using the functional form of Eq. {\ref{x(t)_Eqn}}. The $x(t)$ data is most sensitive to the details of the curling motion, and thus is highly model sensitive, in contrast to $\theta (t)$ and $y(t)$, which are inherently less sensitive to the details of the curling motion

\begin{figure}
\caption{Trajectory data from a rapidly rotating slowly sliding curling rock showing $x(t)$ (top), $y(t)$ (middle) and $\theta(t)$ (bottom). The data for $y(t)$ is fit using Eq. {\ref{y(t)_Eqn}} and that for $\theta(t)$ is fit using Eq. {\ref{q(t)_Eqn}}. The data for $x(t)$ is fit using Eq. {\ref{x(t)_Eqn}} with the i) parameter $\lambda_y=2.268$ from the $y(t)$ fit and $a_T=3.6\times10^{-3} \textrm{m/s}^2$ (solid line) and ii) optimized $\lambda=1.33$ and $a_T=2.4\times10^{-3} \textrm{m/s}^2$ (dashed line).}
\includegraphics[scale=0.8]{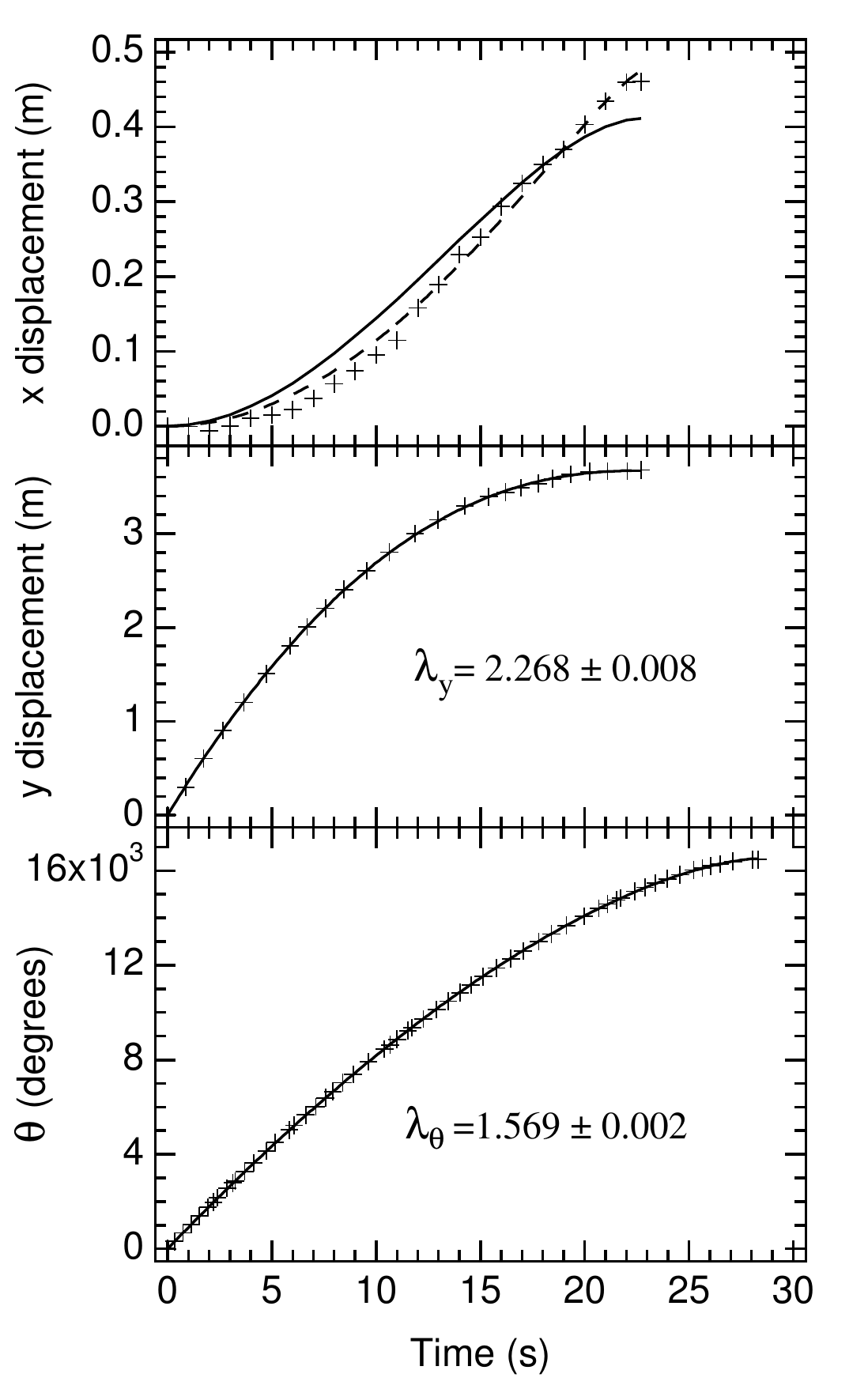}
\label{RRSS_PI_Data1}
\end{figure}

Typical data from a rapidly rotating slowly sliding (RRSS) shot is shown in Fig. {\ref{RRSS_PI_Data1}}. In this particular data set it is observed that translational motion ceases $\Delta t=t_{F_{\theta}}-t_{F_s}=5.5$s prior to rotational motion. We have observed varying amounts for $\Delta t$ in other shots. One contribution to this variation is likely due to the variations in ice pebble heights; a very slowly translating rock can become trapped in a `well' before rotation ceases\footnote{Note that some of the `vibration' that one observes in a sliding curling rock is also due to this slight variation in pebble heights.}. As expected, the rotational motion is well represented by the formula of Eq. {\ref{q(t)_Eqn}}, here with the parameter $\lambda_{\theta}=1.569$, a value similar to that of the comparable purely rotational shots, and also representative of the different RRSS shots we examined. From the examination of a set of RRSS shots, we found a mean value of $\lambda_{\theta}=1.577\pm0.013$.

The $y(t)$ trajectory data for Fig. {\ref{RRSS_PI_Data1}} is fit quite well by the formula of Eq. {\ref{y(t)_Eqn}}, which would indicate that the application of Eq. {\ref{y(t)_Eqn}} in this case is justifiable. The observed value for the friction parameter $\lambda_y\approx2.27$ is comparable to that found in other runs, but we do find significantly more run-to-run variation in $\lambda_y$ than was the case for $\lambda_{\theta}$. We observed values for $\lambda_y$ between 1.7 and 2.4, with no clear correlation of this value with parameters such as $t_{F_S}$ or $y_F$. 

The $x(t)$ trajectory  data shows the characteristic shallow $s$-shape, which was observed for most good curling shots. Although strictly outside its regime of applicability, we fit the $x(t)$ data to the functional form of Eq. {\ref{x(t)_Eqn}}. In the first instance, we fit the data using a fixed $\lambda_y$ parameter- this is a one parameter fit varying only the transverse acceleration $a_T$ and is shown as a solid line in the top curve of Fig. {\ref{RRSS_PI_Data1}}, here with a value $a_T=3.6\times10^{-3} \textrm{m/s}^2$. We did find that in general the $\lambda_y$ values obtained from $y(t)$ data were too large for optimal $x(t)$ fits, so if we allow a fit varying both $\lambda$ and $a_T$ then an improved fit is obtained (dashed line) with $\lambda=1.33$, a value significantly less than the fitted $\lambda_y$ value and with $a_T=2.4\times10^{-3} \textrm{m/s}^2$. Note that, as expected, these values of $a_T$ are much smaller than $\mu g$, which is of the order 0.1m/s$^2$.

For $\lambda_{\theta}=1.569$, Fig. {\ref{RRSS_PI_Data1}}, we find $\phi_{\theta}=-0.7575$ and $\alpha_{\theta}=0.0103$ (\textrm{m/s})$^{-\phi_{\theta} }$, both of which are reasonably close to $\phi_{rot}$ and $\alpha_{rot}$ for pure rotation. Also, using Eq. {\ref{Cy_Eqn}}, we find the value $C_y=1.77$ for this shot, which is a reasonable value.

\begin{figure}
\caption{Comparison of two trajectories for rapidly rotating, slowly sliding curling rocks on pebbled and flooded ice. The shot on pebbled ice travelled for 28.3s and made 45.8 rotations. The shot on flooded ice travelled for 13.3s and made 20.8 total rotations.}
\includegraphics[scale=0.85]{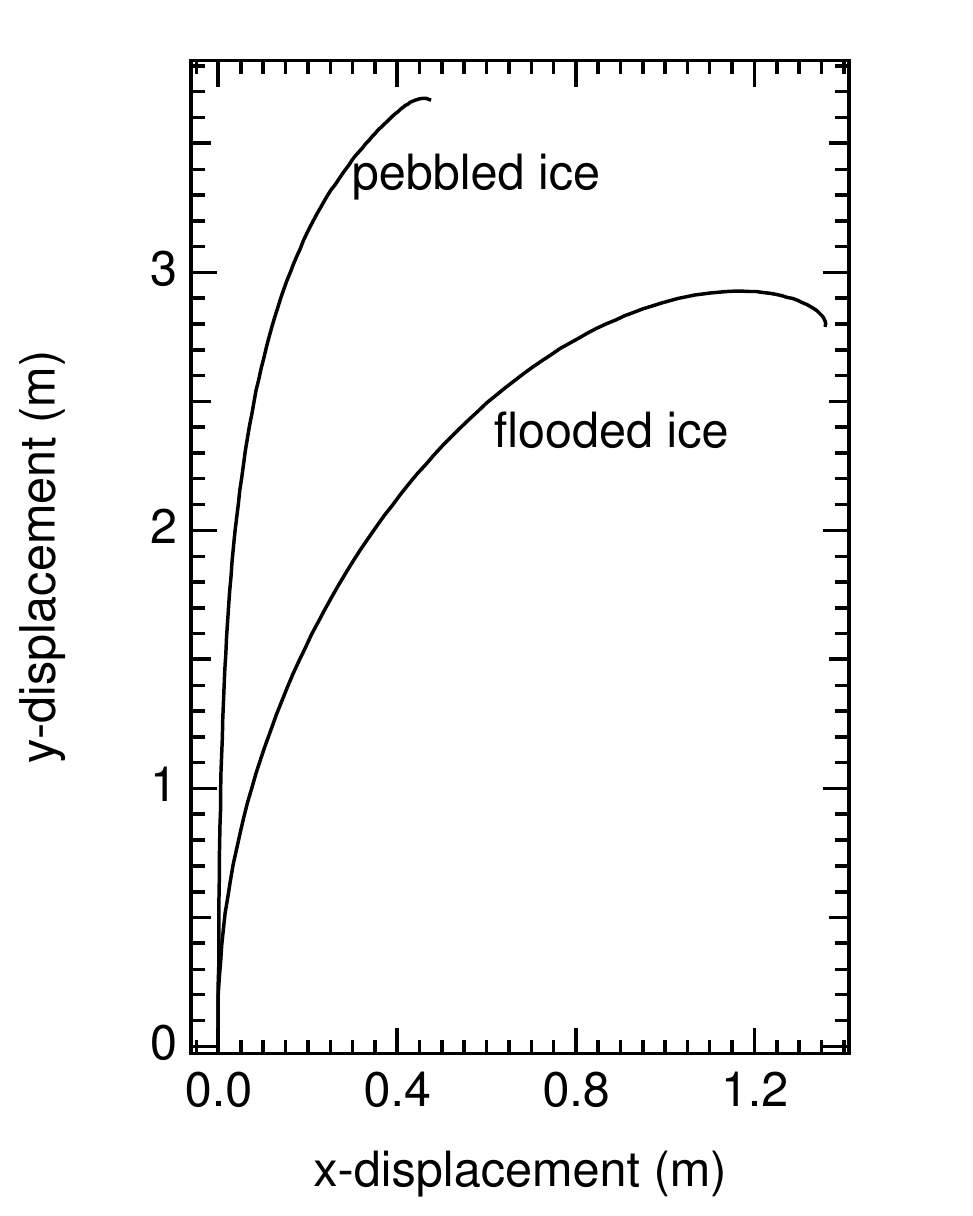}
\label{RRSS_trajectories}
\end{figure}

We also show Figure {\ref{RRSS_trajectories}}, which contrasts two representative curling shot trajectories from a standard curling rock on pebbled ice and flooded ice. Note that although the scaled motions show the same qualitative behavior, the absolute magnitude of the total $x$-motion (i.e. the curl distances are substantially different; much more curl is obtained in the case of flooded ice than from a comparable shot on pebbled ice. The curling shot on flooded ice also shows the interesting feature that the curl can be sufficient to cause the curling rock to have reversed its direction. We have observed this behaviour for  a number of shots on flooded ice, and is a manifestation of the larger curl on the flooded ice surface. Theoretical work investigating such large curl was presented in Ref. \cite{Shegelski:2002}. The shots on flooded ice showed somewhat different values for the shot parameters than on pebbled ice, for example $\overline{\lambda}_{\theta}=1.74\pm0.06$. Analysis of $\lambda_y$ was complicated by the $y$-direction reversal for some shots, which is incompatible with Eq. {\ref{q(t)_Eqn}}. For the subset of short shots, we found $\lambda_y\sim 2.2$. The observed $x$-motions on flooded ice also displayed a best fit using Eq. {\ref{x(t)_Eqn}} for $\lambda$ less than $\lambda_y$, with best fit values of $\lambda$ from 1.4 to 1.8. Most notable was the observation of significantly higher values for the best fit transverse acceleration-- on flooded ice we found values of $a_T$ from $1.5\times10^{-2}\textrm{m/s}^2$ to $3.3\times10^{-2}\textrm{m/s}^2$, roughly ten times higher than shots on pebbled ice.

\subsection{Standard Draw Shot Motion}
As compared to the shots considered in Section {\ref{RRSS_Section}}, these curling shots have a relatively low rotation rate (typically a few rotations) and a longer total trajectory. Also we have found that on pebbled ice the sliding and rotational motions typically cease at the same time, within the resolution of our experiments and in agreement with results previously reported \cite{Voyenli:1985, Farkas:2003}.

\begin{figure}
\caption{Trajectory data for a typical draw shot on pebbled ice. Data is shown for $x(t)$ (top), $y(t)$ (middle) and $\theta(t)$ (bottom). The data for $y(t)$ is fit to Eq. {\ref{y(t)_Eqn}} and for $\theta(t)$ to Eq. {\ref{q(t)_Eqn}} . The data for $x(t)$ is fit to Eq. {\ref{x(t)_Eqn}} using i) the $\lambda_y=1.897$ value from the $y(t)$ fit (solid line) and ii) an optimized $\lambda=1.43$ (dashed line).}
\includegraphics[scale=0.75]{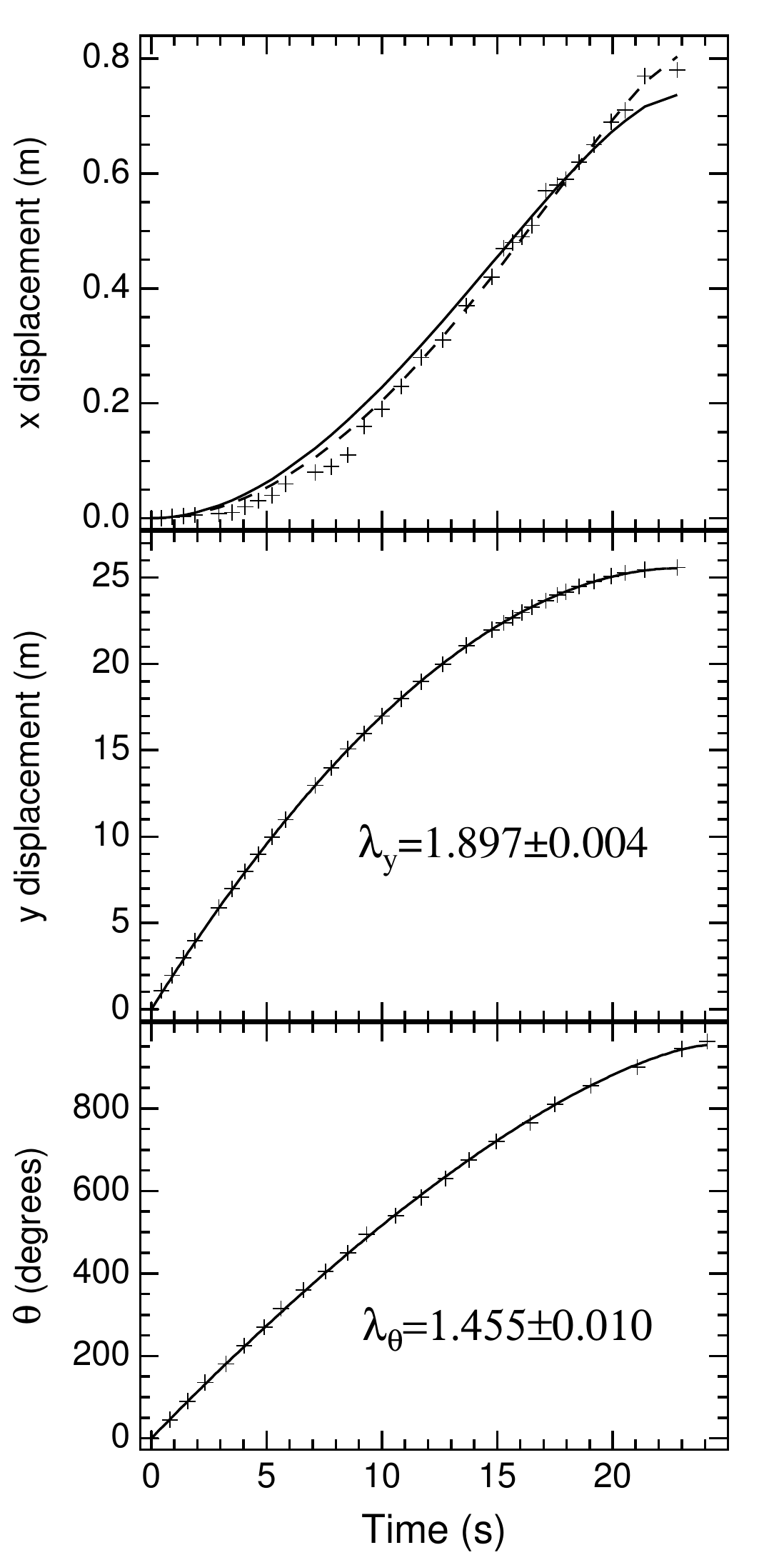}
\label{PI_xyq_Data}
\end{figure}

In Figure {\ref{PI_xyq_Data}}, we show $x(t)$, $y(t)$ and $\theta(t)$ data for a typical draw shot on pebbled ice selected from the large number of shots we recorded. Here the same phenomonological fitting curves were used for $\theta(t)$ and $y(t)$, as shown in Fig. {\ref{PI_xyq_Data}}, the fits to the data are reasonable. The $\theta(t)$ data has noticeably more scatter here, but this is simply due to the smaller number of rotations, in this case 2.7 rotations. Most notable here is that although the same functional form for fitting is used, the extracted parameters $\lambda_{\theta}$ and $\lambda_y$ differ considerably from those of the RRSS shots. In Fig.{\ref{PI_xyq_Data}} $\lambda_{\theta}=1.455$, which is considerably less than observed for the pure rotational data or for the RRSS shots. We did observe some shot-to-shot variation, with $\lambda_{\theta}$ values between 1.20 and 1.50. Averaging over all the observed shots we found a mean value of $\overline{\lambda}_{\theta}=1.32\pm0.10$. For the $\lambda_y$ parameter, we observed a variation in $\lambda_y$ from 1.77 to 1.90, with a mean value of $\overline{\lambda}_y=1.84\pm0.04$. 

For the motion perpendicular to the inital motion $x(t)$, the form of this displacement is qualitatively similar to that observed for the RRSS trajectories. We analyzed $x(t)$ data for a number of curling shots and, in order to try to understand this motion, we attempted to extract a form for the transverse force ${F}_T$ by numerical differentiation of the data. Although such numerical differentiation of noisy data was not definitive, we were led to believe that a roughly constant ${F}_T$ was reasonable for the majority of the curling shot motion. This was the observation that led us to Eq. {\ref{x(t)_Eqn}}. The $x(t)$ data of Fig. {\ref{PI_xyq_Data}} is shown with two fits using Eq. {\ref{x(t)_Eqn}}. The solid curve is obtained using a fixed $\lambda_y=1.897$ and $a_T=5.5\times10^{-3}\textrm{m/s}^2$, and the improved dashed curve fit is obtained if we allow $\lambda$ to vary, here with a value of $\lambda=1.43$ and $a_T=4.6\times10^{-3}\textrm{m/s}^2$. Note that this value for $\lambda$ is quite close to that of $\lambda_{\theta}=1.455$. It is also notable that the observed transverse acceleration $a_T$ is small compared to the deceleration of the curling rock translational motion. For the draw shot of Fig. {\ref{PI_xyq_Data}}, the average deceleration of the rock is $|a_v|\sim  0.09$m/s$^2$, roughly a factor of $20\times$ larger than the transverse acceleration.

Given the observed value of $\overline{\lambda}_y = 1.84 \pm 0.04$, we find $\overline{\phi}_y = -0.19 \pm 0.06$.  For the draw shot of Fig. 5, we find $\alpha_y = 0.0076$ (m/s)$^{-\phi_y}$.  These values are in good agreement with the values of $\phi$ and $\alpha$ calculated above. Also, Eq. {\ref{Cw_Eqn}} leads to $C_{\omega}=2.54$, which is a reasonable value and is in accord with the hypothesis of the direction of friction being shifted due to the thin liquid film.

\begin{figure}
\caption{Comparison between draw shots observed on standard pebbled ice and flooded ice. The pebbled ice shot travelled for 24.1s and rotated 2.7 times, while the flooded ice shot travelled for 18.8s and rotated 2.1 times. }
\includegraphics[scale=1.0]{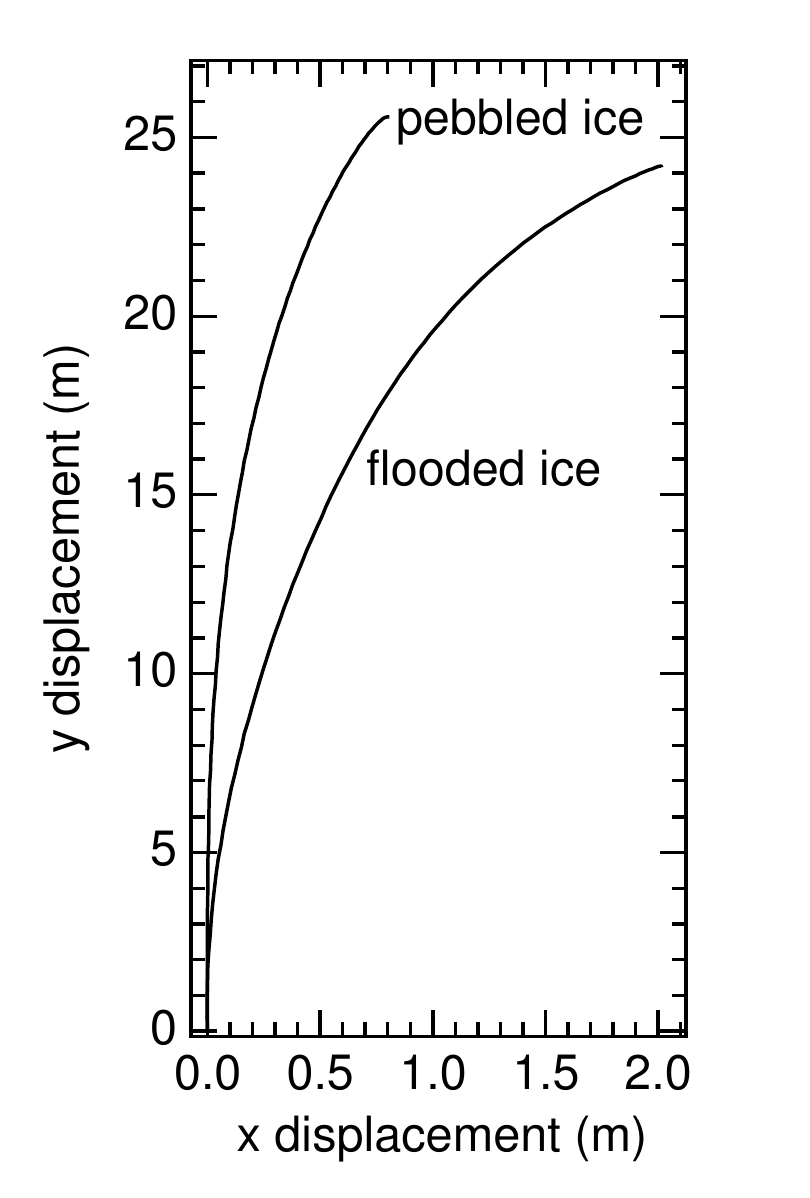}
\label{Trajectory_Data1}
\end{figure}

We have also compared the draw shot trajectory obtained on pebbled ice (Fig. {\ref{PI_xyq_Data}}) with that obtained for a comparable shot on the flooded ice surface. These trajectories are shown in Fig. {\ref{Trajectory_Data1}}. We do observe some differing behaviors for the rock on these different surfaces. In about half of the shots on flooded ice, the $\theta(t)$ curve was qualitatively different from Eq. {\ref{q(t)_Eqn}} and the rotation appeared to cease well before translation. For other $\theta(t)$ curves, an adequate fit was obtained using Eq. {\ref{q(t)_Eqn}} and we found $\lambda_{\theta}$ values from 2.3 to 2.8. The $\lambda_y$ parameter for flooded ice was similar to that on pebbled ice, with an average value $\overline{\lambda}_y=1.74$. The observed $x(t)$ displacements are larger on the flooded ice than on pebbled ice, for comparable curling shots, which is similar to what was found for the RRSS case. The observed transverse accelerations were similar the RRSS shots on flooded ice, we found $\overline{a}_T=2.1\times10^{-2}\textrm{m/s}^2$ and for best fits, $\overline{\lambda}=1.5$. We also note that curling shots we have tried on {\it burned}\/ (i.e. smooth) ice show very little curl and compared to the regular pebbled ice surface, and also shots performed on {\it scraped}\/ (i.e. pebbling removed by a mechanical scraper) ice show extremeley large curl distances.

\subsection{Variation in Number of Turns}
\begin{figure}
\caption{Measured curl distances plotted against the number of turns for a curling rock on pebbled ice. The curling rock release velocities were generally decreased for shots with larger numbers of turns such that the curling rock trajectory finished close to the legal target area of the rink (i.e. roughly fixed trajectory length).}
\includegraphics[scale=0.85]{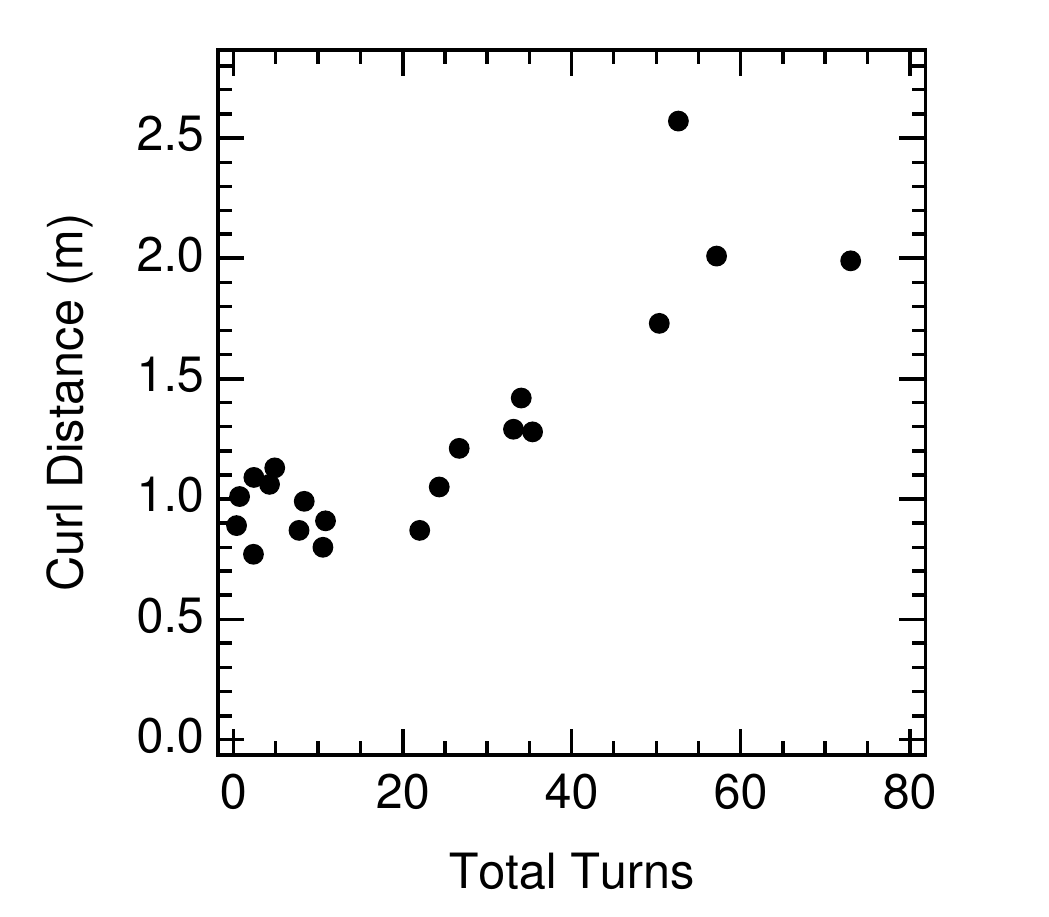}
\label{NT_Data1}
\end{figure}

A further variation from standard curling shots we have examined is the case of increasing initial rotation rate, hence an increased total number of rock rotations $N_T$. In a normal game of curling, these shots ({\it spinners}\/) are occasionally used as a takeout shot to remove stones from the house. To use these shots to place the curling rock inside the legal target area, we found that as the rotation rate was increased and the rotational kinetic energy became large, the rocks had to be released with slower initial velocity in order that they did not pass right through the target area. Thus we attempted to study these shots provided that the trajectories finished near the legal target area (i.e. roughly a fixed $y_{total}$ and not at, say, fixed initial release velocity). It was found that under these conditions, as the number of rock rotations was increased, the total curl distance gradually increased, as shown in Fig. {\ref{NT_Data1}}. The amount of total curl distance did not increase rapidly with $N_T$ however. Rocks that were thrown for a total of 50-80 total turns would curl a distance roughly twice that of more standard shots having 2-10 turns. We did not observe any decrease in total $x$ displacement as $N_T$ increased, in contrast with some earlier work \cite{Penner:2001}. 

\begin{figure}
\caption{Trajectory data for a `spinner' shot made on pebbled ice. The plots show $x(t)$ (top), $y(t)$ (middle), and $\theta(t)$ (bottom). The fit to $y(t)$ data using Eq. {\ref{y(t)_Eqn}} is notably poorer for these spinner shots. The data for $x(t)$ is fit to Eq. {\ref{x(t)_Eqn}} using i) the $\lambda_y=2.30$ value from the $y(t)$ fit (solid line) and ii) an optimized $\lambda=1.57$ (dashed line). }
\includegraphics[scale=0.73]{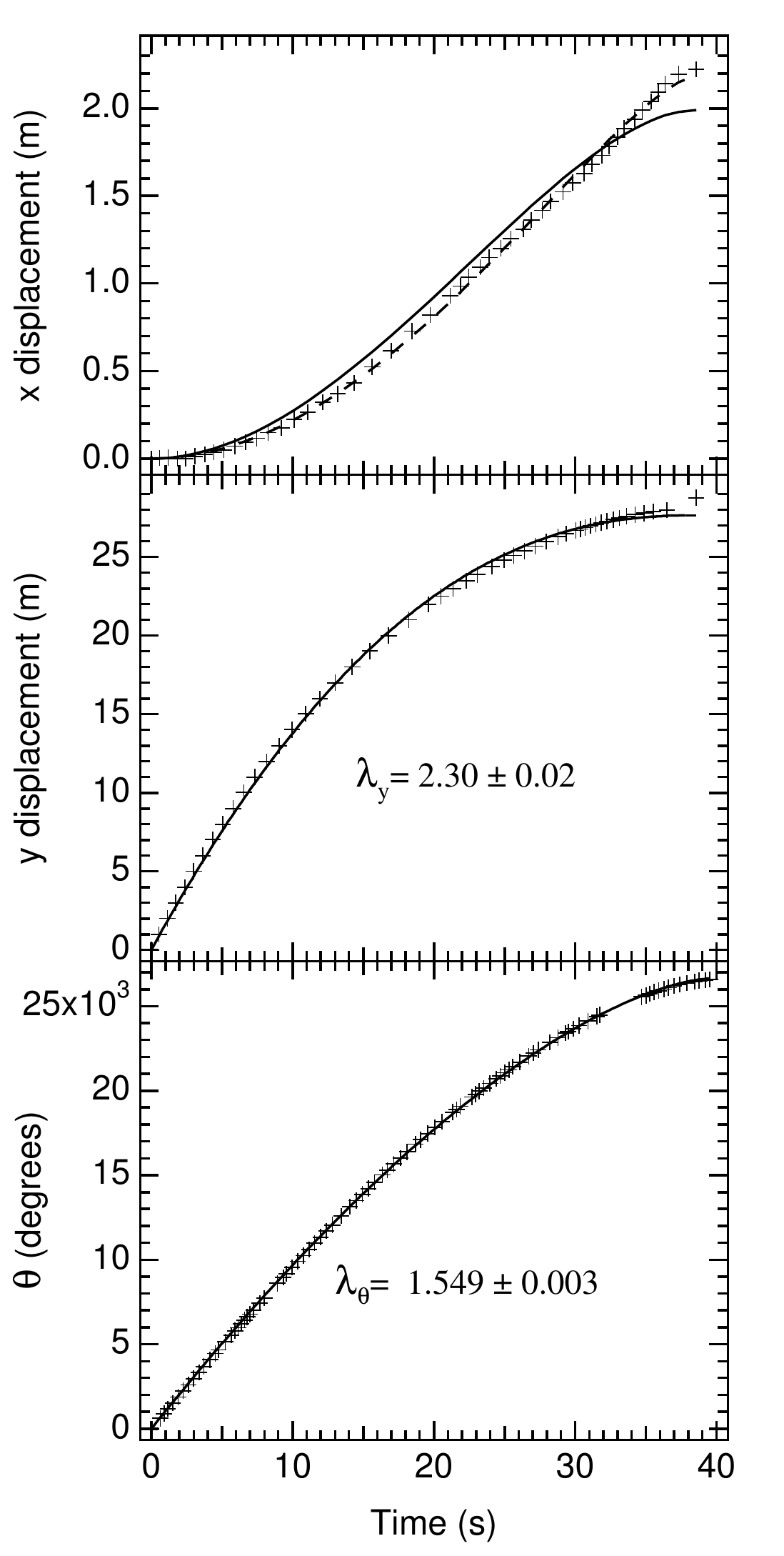}
\label{NT_x_y_q_data}
\end{figure}

Figure {\ref{NT_x_y_q_data}} shows $x(t)$, $y(t)$, and $\theta(t)$ data from a typical spinner shot on pebbled ice. For these shots having large $N_T$ values, we found that the phenomenological fitting function Eq. {\ref{y(t)_Eqn}} for $y(t)$ generally did not properly represent the observed trajectories as well as for the cases previously discussed. The best fit curve systematically undershoots then overshoots the data. The fit to the $\theta(t)$ data is quite good, with the best fit $\lambda_{\theta}$ value somewhat lower than the pure rotation values. The fitted curve to the $x(t)$ data is shown for both fixed $\lambda_y=2.30$ and optimized $\lambda=1.57$.

\section{Conclusions and Outlook}
We have collected data from a large number of curling shots using a wide range of rotational and translational velocities as well as having studied motions on two different ice surfaces. Since a complete first principles approach to modeling the observed trajectories is still lacking, we have used a semi-phenomenological approach to describe the trajectories. We have shown that the observed trajectories of curling rocks can be very accurately represented through the use of Equations {\ref{q(t)_Eqn}}, {\ref{y(t)_Eqn}}, and {\ref{x(t)_Eqn}} and that these trajectories can be represented using a small number of parameters. 

The experimental results reported in this paper show unequivocally that dry friction, i.e. constant $\mu$, cannot account for any of the observed motions.  The granular model proposed by Denny, based on constant $\mu$, cannot account for the experimental results reported here. The granular model \cite{Denny:2002} applied to a draw shot predicts $\lambda_y=2$ and $\lambda_{\theta}=1.1$, which differs from the observed values of $\lambda_y=1.84\pm0.04$ and $\lambda_{\theta}=1.32\pm0.10$.

The present study has found a number of pieces of evidence supporting the existence of a thin liquid film between the contact points of the annulus and the underlying solid ice, i.e. wet friction. The observed values for the exponent $\phi$ are inconsistent with `dry friction' models, and we also find that the most facile wet friction models using $\phi=-1/2$ cannot account for the observed motions. We have found evidence that the direction of the frictional force acting upon a segment of the curling rock is not relative to that of the fixed ice surface but instead relative to the moving thin liquid film. This is supported by the observation that rotational and translational motion need not cease simultaneously, such as in the case of rapidly-rotating--slowly sliding motion on pebbled ice. This hypothesis is supported by the finding that the parameters $C_{\omega}$ and $C_y$ have values of order 1 but not equal to 1. Further supporting evidence is the observation that current theoretical models cannot account for the magnitude of the observed curl-- the observed curl distances being nearly a factor of two larger. In our view, it is the redirection of the frictional force to be relative to the  thin film which causes larger than expected curl distances. This is also related to the finding here that the curl distance is only very weakly related to the initial rotational rate-- we do find that $x_F$ increases with the number of rotations, but only very weakly and in such a way that cannot be explained by a dry friction model, yet is readily understood in terms of a thin liquid film. Due to the rapidity of rotation for these shots, the lubrication of the thin film results only a marginally larger transverse force than a rock that is rotating more slowly.

The fit of the semi-phenomenological treatment to the lateral displacement $x(t)$ of the rock as a function of time was surprisingly good, especially given the simplicity of the functional form used for $\vec{F}_T$. The present study has found that the curl trajectory $x(t)$ can be effectively modeled by a constant transverse force. While we do not expect that the transverse force is truly constant, we do believe that the actual force $\vec{F}_T(t)$ will be roughly constant through the majority of the motion.

Some rather unexpected results were found as well.  Curling rocks curl much more on {\em flooded} ice than they do on {\em pebbled} ice.  This was found for draw shots and also for rapid-rotation---slow-sliding shots. We also found that shots with a large number of rotations (spinners) {\em curl more} than standard draw shots by a factor of 2 or so.

It is hoped that future work will be able to model the interaction between the curling rock and the thin liquid film in a more complete manner, and show in detail how the frictional force acting upon the rock deviates in direction from what would be expected for dry friction.

\section*{Acknowledgments}
We would like to thank the staff at the Prince George Golf and Curling Club for their assistance, particularly Murray Kutyn and John and Raymond Wercholuk. We also thank the following students for their assistance in collecting and collating the data: Annette Brown, Adam Gerber, Dino Gigliotti and Erik Kozijn. Technical and financial assistance from Francine Poisson and Keith Cobelt of CanCurl is gratefully acknowledged. Financial support from the Natural Sciences and Engineering Research Council of Canada is also gratefully acknowledged.

\newpage

\end{document}